\documentclass[journal,10pt,twoside,twocolumn]{IEEEtran}
\IEEEoverridecommandlockouts

\IEEEoverridecommandlockouts

 \usepackage{algorithm}
 \usepackage{algorithmic}
\usepackage{amsmath,amsfonts} 
\usepackage{amssymb, amsthm}
\usepackage{graphicx}
\usepackage{subfigure}
\usepackage{booktabs}
\usepackage{multirow}
\usepackage{cite}
\usepackage{upgreek}
\usepackage{color}
\usepackage{microtype}
\usepackage{balance}
\usepackage{array}
\usepackage{hyperref}

\renewcommand{\eqref}[1]{(\ref{#1})}
\newcommand{\secref}[1]{\mbox{Section~\ref{#1}}}

\newcommand{\figref}[1]{\mbox{Figure~\ref{#1}}}
\newcommand{\tblref}[1]{\mbox{Table~\ref{#1}}}
\newcommand{\algref}[1]{\mbox{Algorithm~\ref{#1}}}

\newcommand{\Ex}[1]{\ensuremath{\mathbb{E}\!\left[#1\right]}}

\renewcommand{\mid}{\ensuremath{\,|\,}}

\DeclareMathOperator*{\argmin}{arg\;min}

\newcommand{\abs}[1]{\ensuremath{\left|#1\right|}}

\newcommand{\ceil}[1]{\ensuremath{\left\lceil#1\right\rceil}}

\newcommand{\setA}{\ensuremath{\mathcal{A}}}

\newcommand{\setC}{\ensuremath{\mathcal{C}}}

\newcommand{\setO}{\ensuremath{\mathcal{O}}}

\newcommand{\bma}{\ensuremath{\mathbf{a}}}

\newcommand{\bmh}{\ensuremath{\mathbf{h}}}

\newcommand{\bmn}{\ensuremath{\mathbf{n}}}

\newcommand{\bmr}{\ensuremath{\mathbf{r}}}
\newcommand{\bms}{\ensuremath{\mathbf{s}}}

\newcommand{\bmy}{\ensuremath{\mathbf{y}}}
\newcommand{\bmz}{\ensuremath{\mathbf{z}}}

\newcommand{\bA}{\ensuremath{\mathbf{A}}}

\newcommand{\bG}{\ensuremath{\mathbf{G}}}
\newcommand{\bH}{\ensuremath{\mathbf{H}}}
\newcommand{\bI}{\ensuremath{\mathbf{I}}}

\newcommand{\bZero}{\ensuremath{\mathbf{0}}}

\newcommand{\MT}{{\ensuremath{U}}}
\newcommand{\MR}{{\ensuremath{B}}}

\newcommand{\No}{\ensuremath{N_{\mathrm{0}}}}

\newcommand{\Eb}{{\ensuremath{E_{\mathrm{b}}}}}



\newcommand{\rev}[1]{\textcolor{black}{#1}}

\begin{document}
\title{High-Throughput Data Detection for Massive MU-MIMO-OFDM using Coordinate Descent }
\author{
\IEEEauthorblockN{Michael Wu, Chris Dick, Joseph R.\ Cavallaro, and Christoph Studer} 
\thanks{M.~Wu and J.~R.~Cavallaro are with the Department of ECE, Rice University, Houston, TX; e-mail: \{mbw2,\,cavallar\}@rice.edu} 
\thanks{M.~Wu and C.~Dick are with Xilinx Inc., San Jose, CA; e-mail: \{miwu,\,chris.dick\}@xilinx.com}
\thanks{C.~Studer  is with the  School~of ECE, Cornell University, Ithaca, NY; e-mail: studer@cornell.edu} 
\thanks{A short version of this paper for a single-carrier frequency-division multiple access (SC-FDMA) massive MU-MIMO systems has been presented at the  IEEE International Symposium on Circuits and Systems (ISCAS)~\cite{Wu2016}.}
\thanks{A simple massive MU-MIMO system simulator to experiment with OCD is available on GitHub: \url{https://github.com/VIP-Group/OCD_sim}}
}

\maketitle

\begin{abstract}
Data detection in massive multi-user (MU) multiple-input multiple-output (MIMO) wireless systems is among the most critical tasks due to the excessively high implementation complexity.
In this paper, we propose a novel, equalization-based soft-output data-detection algorithm and corresponding reference FPGA designs for wideband massive MU-MIMO systems that use orthogonal frequency-division multiplexing (OFDM).
Our data-detection algorithm performs approximate minimum mean-square error (MMSE) or box-constrained equalization using coordinate descent.
We deploy a variety of algorithm-level optimizations that enable near-optimal error-rate performance at low implementation complexity, even for systems with hundreds of base-station (BS) antennas and thousands of subcarriers. 
We design a parallel VLSI architecture that uses pipeline interleaving and can be parametrized at design time to support various antenna configurations.
\rev{We develop reference FPGA designs for massive MU-MIMO-OFDM systems and  provide an extensive comparison to existing designs in terms of implementation complexity, throughput, and error-rate performance.}
For a 128 BS antenna, 8 user  massive MU-MIMO-OFDM system, our FPGA design outperforms the next-best implementation by more than~$\bf 2.6\boldsymbol\times$ in terms of throughput per FPGA look-up tables.
\end{abstract}


%


\section{Introduction}
\label{sec:intro}
\IEEEPARstart{M}{assive} multi-user (MU) multiple-input multiple-output (MIMO) technology promises significant improvements in terms of spectral efficiency, coverage, and range compared to traditional, small-scale MIMO~\cite{Marzetta2010,Rusek2012,hoydis2013massive,larsson2014massive}. In fact, massive MU-MIMO is commonly believed to be one of the key technologies for future fifth-generation (5G) wireless systems~\cite{andrews2014will}. 
The idea underlying this technology is to equip the base-station (BS) with hundreds of antenna elements while communicating with tens of user terminals concurrently and within the same time-frequency resource. 
\rev{However, the large dimensionality of the data detection problem faced in the uplink (where users communicate to the BS), results in excessively high implementation complexity at the BS~(see, e.g., \cite{Wu2014} and the references therein).}
Hence, to reduce the implementation costs while enabling throughputs in the Gb/s regime for practical wideband massive MU-MIMO systems with hundreds of antenna elements and thousands of subcarriers, novel algorithms and dedicated hardware implementations on field-programmable gate arrays (FPGAs) or application specific integrated circuits (ASICs) are  necessary. 

During recent years, various data-detection algorithms~\cite{Prabhu2013,Hu2014} and dedicated hardware implementations have been proposed for massive MU-MIMO systems~\cite{Wu2014,Yin2015,YWWDCS14b,CGS2016a,WZXXY2016}.
\rev{All of the existing hardware implementations, however, are either unable to achieve the high throughputs offered by future wideband massive MU-MIMO systems  \cite{Wu2014,CGS2016a,WZXXY2016}, or exhibit excessive hardware complexity~\cite{YWWDCS14b}.}
\rev{Furthermore, the hardware implementations in \cite{Wu2014,YWWDCS14b} only support single-carrier frequency-division multiple-access (SC-FDMA). As demonstrated in~\cite{tunali15}, however, orthogonal frequency-division multiplexing (OFDM) enables (often significantly) less complex  baseband processing\footnote{\rev{SC-FDMA typically generates baseband signals with a lower dynamic range, but the receiver must perform an additional frequency-to-time conversion (compared to OFDM). This additional conversion step requires one to separate equalization (that is usually carried out in the frequency domain per subcarrier) and data detection (that must be carried out in the time domain). This separation prevents the use of powerful, non-linear equalization methods~\cite{wuiterativeLTE}, such as the box-constrained detector proposed in this paper. OFDM, in contrast, causes a slightly higher dynamic range, but requires only one time-to-frequency conversion and enables non-linear data-detection methods that operate directly in the frequency domain on a per-subcarrier basis~\cite{tunali15}.}}, which may be a critical design factor for wideband massive MU-MIMO systems with hundreds of BS antennas and thousands of subcarriers.}

\subsection{Contributions}
We propose a new, low-complexity soft-output data-detection algorithm and a corresponding high-throughput FPGA design for massive MU-MIMO wireless systems that use OFDM. 
\rev{Our algorithm, referred to as \underline{o}ptimized \underline{c}oordinate \underline{d}escent (OCD), performs approximate minimum mean-square error (MMSE) or box-constrained equalization, which enables near maximum-likelihood (ML) soft-output data detection performance in massive MU-MIMO systems with a large BS-to-user-antenna ratio.}
We develop a corresponding high-throughput VLSI architecture with a deep and interleaved pipeline, which can be parametrized at design time to support various BS and user antenna configurations. 
\rev{The algorithmic regularity of OCD and the fact that preprocessing can be implemented at minimum hardware overhead enables high-throughput VLSI designs that require lower complexity than state-of-the-art designs, even for systems with hundreds of BS antennas and thousands of subcarriers.}
To demonstrate the advantages of OCD compared to existing massive MU-MIMO data-detector designs in terms of throughput, hardware complexity, and error-rate performance, we provide implementation results on a Xilinx Virtex-7 FPGA.

\subsection{Notation}

Boldface lowercase and boldface uppercase letters stand for column vectors and matrices, respectively. For a matrix~\bA, we denote its hermitian transpose by $\bA^H$.
We use  $a_{k,\ell}$ for the entry in the $k$th row and $\ell$th column of the matrix $\bA$; the $k$th entry of a column vector $\bma$ is~denoted by $a_k=[\bma]_k$. The $\ell_2$-norm of a vector $\bma$ is defined as~$\|\bma\|_2=\sqrt{\sum_{k}|a_k|^2}$.  \rev{The real part of a complex number $a$ is $\Re\{a\}$. Sets are denoted by uppercase calligraphic letters; the cardinality of a set~$\setA$ is~$|\setA|$. The expectation operator is designated by $\Ex{\cdot}$.}

\subsection{Paper Outline}
The rest of the paper is organized as follows. 
\secref{sec:system} introduces the massive MU-MIMO-OFDM system model and describes data detection using MMSE and box-constrained equalization.
\secref{sec:algo} details our OCD algorithm and shows error-rate simulation results.
\secref{sec:impl} and \secref{sec:implementation}  describe our VLSI architecture and 
shows FPGA implementation results, respectively. 
We conclude in \secref{sec:conclusions}.

\section{System Model and Data Detection}
\label{sec:system}
This section introduces the considered OFDM-based uplink model and summarizes efficient methods for linear MMSE and box-constrained  soft-output data detection. 
\subsection{OFDM-based Uplink System Model}

We consider a massive MU-MIMO-OFDM uplink system, where~$\MT$ single-antenna user terminals send data {simultaneously} to a BS with~$\MR\gg\MT$ antennas over $W$ subcarriers.  Each user $i=1,\ldots,\MT$ encodes its own bit stream (using a forward error-correction scheme) and maps the generated coded bits onto constellation points in a finite set $\setO$ (e.g., \mbox{64-QAM} using a Gray mapping rule), with unit average transmit power, i.e., \mbox{$\Ex{|s|^2}=1$} with $s\in\setO$, and \rev{$Q=\log_2|\setO|$} bits per constellation point. The resulting $W$ frequency-domain symbols $\{s^{(i)}_1,\ldots,s^{(i)}_W\}$ are then transformed into the time domain (TD) using an inverse discrete Fourier transform (DFT)~\cite{OFDM2004}. 
After prepending the cyclic prefix, all users transmit their TD signals  over the frequency-selective wireless channel at the same time.

After removing the cyclic prefixes, the TD signals received at each BS antenna are transformed back to the FD using a DFT. For the sake of simplicity, we assume a sufficiently long cyclic prefix,  perfect synchronization, and that perfect channel-state information (CSI) has been acquired via pilot-based training.\footnote{These assumptions are common in the MIMO-OFDM literature \cite{OFDM2004}.} Under these assumptions, the FD input-output relation on the $w$th  subcarrier is commonly  modeled as~\cite{gesbert2003theory}
\begin{align} \label{eq:OFDMsystem}
\bmy_w = \bH_w\bms_w +\bmn_w, 
\end{align}
where $\bmy_w\in\mathbb{C}^\MR$ is the associated received FD vector, $\bH_{w}\in\mathbb{C}^{\MR\times\MT}$ is the channel matrix, $\bms_w\in\setO^\MT$ contains the symbols transmitted by all $\MT$ users, i.e., $[\bms_w]_i=s^{(i)}_w$ refers to the symbol transmitted by user $i$ over subcarrier $w$, and $\bmn_w\in\mathbb{C}^\MT$ models thermal noise as  i.i.d.\  complex circularly-symmetric Gaussian vector with variance $N_0$ per complex entry.

\subsection{Equalization-based Data Detection}
For the model in \eqref{eq:OFDMsystem}, optimal data detection in terms of minimizing the  symbol error-rate is accomplished by solving the maximum-likelihood (ML) problem~\cite{Paulraj2008}
\begin{align} \label{eq:MLproblem}
 \tilde{\bms}^\text{ML}_w = \argmin_{\bmz\in\setO^\MT} \|\bmy_w-\bH_w\bmz\|_2^2.
\end{align}
Unfortunately, solving \eqref{eq:MLproblem} exactly for massive MU-MIMO systems quickly results in prohibitive complexity, even with the best-known sphere-decoding algorithms~\cite{seethaler2011complexity}. 
Equalization-based data detection algorithms \cite{Paulraj2008} enable one to find approximate solutions to the ML problem at low computational  complexity. 
Virtually all linear as well as non-linear equalization methods relax the finite-alphabet constraint $\bmz\in\setO^\MT$ in \eqref{eq:MLproblem}, which enables the efficient computation of an estimate $\tilde{\bms}$ that is (hopefully) close to the ML solution. The estimate $\tilde{\bms}$ can then either be sliced element-wise onto the nearest constellation point in $\setO$ as follows:
\begin{align} \label{eq:slicing}
\hat{s}_i = \argmin_{z\in\setO} |[\tilde{\bms}]_i-z|, \quad i=1,\ldots,U,
\end{align}
which is known as hard-output data detection, or be used to compute reliability information for each transmitted bit in the form of log-likelihood ratio (LLR) values (see \secref{sec:soft}), which is known as soft-output data detection~\cite{seethaler2004efficient,Studer2011}.

\subsection{Linear MMSE Equalization}
\label{sec:exactsoft}
The most common equalization-based data detection algorithm is linear MMSE data detection~\cite{Paulraj2008,seethaler2004efficient}. This method was shown to enable FPGA and ASIC designs that are able to achieve high throughput in massive MU-MIMO systems~\cite{Wu2014}. \rev{Furthermore, for systems with large BS-to-user antenna ratios~$\delta=\MR/\MT$ (e.g., two or larger), linear detectors are able to achieve near-ML error-rate performance~\cite{Rusek2012,hoydis2013massive,larsson2014massive}.}

The key idea of MMSE data detection is to relax the constraint~$\bmz\in\setO^\MT$ in the ML problem~\eqref{eq:MLproblem} to the $\MT$-dimensional complex space $\bmz\in\mathbb{C}^\MT$, and to include a quadratic penalty function. In particular, MMSE equalization solves the following regularized least-squares problem~\cite{Yincg,Yin2015}:
\begin{align} \label{eq:MMSEproblem}
 \tilde{\bms}^\text{MMSE}_w = \argmin_{\bmz\in\mathbb{C}^\MT} \|\bmy_w-\bH_w\bmz\|_2^2 + \No\|\bmz\|^2_2.
\end{align}
Since the objective function in \eqref{eq:MMSEproblem} is quadratic in $\bmz$, the MMSE equalization problem has a closed-form solution. 

An explicit solution to \eqref{eq:MMSEproblem} can be computed as follows. First, compute the regularized Gram matrix $\bA_w= \bG_w+N_0\bI_\MT$ with  $\bG_w = \bH_w^H\bH_w$ and the matched filter vector $\tilde{\bms}^{\text{MF}}_w = \bH_w^H\bmy_w$. Then, the MMSE estimate in~\eqref{eq:MMSEproblem} is computed as
\begin{align} 
\rev{\tilde{\bms}_w^\text{MMSE} = \bA_w^{-1}\tilde{\bms}^{\text{MF}}_w.}
\end{align}
While this closed-form approach was shown to be efficient for traditional, small-scale MIMO systems (e.g., \rev{with four antennas} at both ends of the wireless link)~\cite{Studer2011}, computing the regularized Gram matrix $\bA_w$ and its inverse~$\bA_w^{-1}$ quickly results in prohibitive complexity in massive MU-MIMO systems with hundreds of BS antennas~\cite{YWWDCS14b}.
In \secref{sec:algo}, we present a computationally-efficient equalization algorithm that directly solves~\eqref{eq:MMSEproblem} in a hardware efficient way, which avoids expensive calculations such as the computation of the  regularized Gram matrix~$\bA_w$ and its inverse~$\bA_w^{-1}$.

\subsection{Non-Linear Box-Constrained (BOX) Equalization}
\label{sec:boxdetectorexact}
While linear equalization methods are the most common approach in the MIMO literature, a few non-linear equalizers have recently emerged and were shown to outperform linear methods in terms of error-rate performance~\cite{JMS2016conf}.
A promising non-linear equalization method, referred to as box-constrained equalization (short BOX equalization) \cite{tan2001constrained,yener2002cdma,TAXH2015}, relaxes the constraint \mbox{$\bmz\in\setO^\MT$} to the convex polytope~$\setC_\setO$ around the constellation set~$\setO$, which is  formally defined as follows:
\begin{align} 
\rev{\setC_\setO = \left\{\sum_{i=1}^{\abs{\setO}}\alpha_i s_i \mid (\alpha_i\geq0, \alpha_i \in\mathbb{R},\forall i) \wedge \sum_{i=1}^{\abs{\setO}}\alpha_i=1 \right\}\!.}
\end{align}
For example, the convex polytope $\setC_\text{QPSK}$ for QPSK with\footnote{We note that this constellation is not normalized to unit expected power.}
\begin{align} 
\rev{\setO=\{+1+j,+1-j,-1+j,-1-j\}}
\end{align} 
is given by $\setC_\text{QPSK}=\{x_R+jx_I: x_R,x_I\in[-1,+1]\}$ with $j^2=-1$; this is simply a box with radius~$1$ around the square constellation (thus the name BOX equalization). \rev{For higher-order QAM alphabets, such as 16-QAM or 64-QAM, we have $\setC_\setO=\{x_R+jx_I: x_R,x_I\in[-\alpha,+\alpha]\}$, where $\alpha=\max_{a\in\setO}\Re\{a\}$ is the radius of the tightest box around the square constellation.}

BOX equalization solves the following relaxed version of the ML problem in \eqref{eq:MLproblem}:
\begin{align} \label{eq:BOXproblem}
 \tilde{\bms}^\text{BOX}_w = \argmin_{\bmz\in\setC_\setO^\MT} \|\bmy_w-\bH_w\bmz\|_2^2.
\end{align}
Since this equalization problem \eqref{eq:BOXproblem} is convex, it can be solved exactly using well-established numerical methods from convex optimization~\cite{BV04}. 
\rev{Furthermore, as shown recently in \cite{TAXH2015,JMS2016conf}, the BOX equalizer exhibits near-ML error-rate performance in the large-antenna limit, where we fix the BS-to-user antenna ratio \mbox{$\delta=\MR/\MT$} so that $\delta>1/2$ and by letting $B\to\infty$. In addition, the BOX equalizer does only need knowledge of the transmit constellation $\setO$ but not of the noise variance $\No$, which is in stark contrast to the MMSE equalizer.}

Unfortunately, solving \eqref{eq:BOXproblem} exactly with conventional interior-point methods results in prohibitive complexity and requires high numerical precision, which prevents efficient hardware designs that use finite precision (fixed-point) arithmetic.  
In order to solve \eqref{eq:BOXproblem} at low complexity and in a hardware efficient manner, we propose a new algorithm in \secref{sec:algo}.

\subsection{Soft-Output Data Detection}
\label{sec:soft}
From MMSE and BOX equalization, hard-output estimates can easily be obtained by element-wise slicing of the entries of $\tilde{\bms}_w^\text{MMSE}$ and $\tilde{\bms}_w^\text{BOX}$  onto the nearest constellation point as in~\eqref{eq:slicing}, respectively. 
In systems that use forward error-correction, however, one is generally interested in soft-output detection~\cite{caire1998bit}. 
From MMSE equalization where $\tilde{\bms}_w=\tilde{\bms}_w^\text{MMSE}$, LLR values are typically computed via the max-log approximation~\cite{Studer2011}
\begin{align}\label{eq:exactinversellr}
L_{w,i,b} = \rho_{w,i}\!\left( \min_{a\in\setO_b^0}\left| \frac{[\tilde{\bms}_w]_i}{\mu_{w,i}} - a\right|^2\! \!\!-\! \min_{a\in\setO_b^1}\left| \frac{[\tilde{\bms}_w]_i}{\mu_{w,i}} - a\right|^2 \right)\!,
\end{align}
where the sets $\setO_b^0$ and $\setO_b^1$ contain the constellation symbols for which the $b$th bit is~$0$ and~$1$, respectively. 
For explicit MMSE detection, i.e., the approach  discussed in \secref{sec:exactsoft} that computes~$\bA_w^{-1}$, the post-equalization signal-to-noise-and-interference-ratio (SINR)~$\rho_{w,i}$ and the channel gain $\mu_{w,i}$ can be calculated exactly and in the following efficient way~\cite{Studer2011}. 
The SINR is calculated as $\rho_{w,i}=\mu_{w,i}/(1-\mu_{w,i})$ and the channel gain is $\mu_{w,i}= [\bA_w]_i^H [\bG_w]_i$, where $[\bA_w]_i$ is the $i$th row of $\bA_w^{-1}$  and~$[\bG_w]_i$ is the~$i$th column of $\bG_w$. 

However, for BOX equalization in \secref{sec:boxdetectorexact}, as well as for data detection algorithms that implicitly solve the MMSE detection problem \eqref{eq:MMSEproblem}, no efficient methods that exactly compute the SINR~$\rho_{w,i}$ are known---this prevents a straightforward computation of the LLR values in  \eqref{eq:exactinversellr}. In \secref{sec:llrapprox}, we propose an approximate way to compute~$\rho_{w,i}$ and~$\mu_{w,i}$, which enables us to generate approximate LLR values for such linear and non-linear equalizers.

\section{Fast Equalization via Coordinate Descent}
\label{sec:algo}
While the solution to the implicit MMSE problem \eqref{eq:MMSEproblem} can be computed (exactly or approximately) at moderate complexity using iterative conjugate gradient (CG) or Gauss-Seidel (GS) methods, see, e.g.,~\cite{Yincg,Hu2014,WZXXY2016}, corresponding VLSI designs~\cite{Yin2015,WZXXY2016} are unable to achieve high throughput, mainly due to a fairly complex algorithm structure, stringent data dependencies, or the need for high arithmetic precision.
We next propose an alternative method to solve both the  MMSE equalization~\eqref{eq:MMSEproblem} and BOX equalizaton \eqref{eq:BOXproblem} problems at low complexity and in a hardware friendly way.

\subsection{Coordinate Descent (CD)}
\label{sec:CDalgo}
Coordinate descent (CD)~\cite{wright2015coordinate} is a well-established iterative framework to exactly or approximately solve a large number of  convex optimization problems using a series of  simple, coordinate-wise updates. 
We first define  the following function:
\begin{align} \label{eq:function}
f(z_1,\ldots,z_U)= f(\bmz) = \|\bmy_w-\bH_w\bmz\|^2_2 + g(\bmz),
\end{align}
where $g(\bmz)$ is a convex regularizer. It is now important to realize that both  equalization problems \eqref{eq:MMSEproblem} and~\eqref{eq:BOXproblem} are special cases when minimizing \eqref{eq:function}. In fact, by setting $g^\text{MMSE}(\bmz)=\No\|\bmz\|_2^2$, minimizing \eqref{eq:function} is equivalent to solving the MMSE equalization problem \eqref{eq:MMSEproblem}. By setting $g^\text{BOX}(\bmz)=\chi(\bmz\in\setC_\setO)$, where \mbox{$\chi(\bmz\in\setC_\setO)$} denotes the characteristic function that is zero if $\bmz\in\setC_\setO$ and infinity otherwise, minimizing~\eqref{eq:function} is equivalent to solving the BOX equalization problem \eqref{eq:BOXproblem}. 
CD-based equalization simply minimizes the function $f(z_1,\ldots,z_U)$ in \eqref{eq:function} sequentially for each variable (or coordinate) $z_u$, $u=1,\ldots,U$, in a round-robin fashion.\footnote{The performance of CD can often be improved by using a carefully-selected variable-update order~\cite{wright2015coordinate}; our own experiments have shown that for MMSE and BOX equalization, a simple round-robin update scheme performs well and is easier to implement.} For more details on CD, see~\cite{wright2015coordinate,Gordon12} and the references therein. We next detail the CD algorithms for MMSE and BOX equalization.

\subsubsection{CD-based MMSE Equalization} 
Assume we want to find the  $u$th optimum value $z_u$ for the MMSE equalization problem~\eqref{eq:MMSEproblem}, i.e., we seek to compute the solution to 
\begin{align} 
\rev{\hat{z}_u = \argmin_{z_u\in\mathbb{C}} \|\bmy_w-\bH_w\bmz\|^2_2 + \No\|\bmz\|_2^2,}
\end{align}
where we hold all other values $z_j$, $\forall j\neq u$, fixed. 
Since this is a quadratic problem, we can solve it in closed form by setting the gradient of the function \eqref{eq:function} with respect to the $u$th component to zero:
\begin{align} \label{eq:gradient} 
 0 = \nabla_u f(\bmz) = \bmh^H_u (\bH\bmz - \bmy) + \No z_u.
 \end{align}
\rev{By decomposing $ \bH\bmz = \bmh_uz_u + \sum_{j\neq u} \bmh_j z_j$, we can solve~\eqref{eq:gradient} for $z_u$ to obtain the following closed-form expression:}
\begin{align} \label{eq:CDrule} 
\hat{z}_u = \frac{1}{\|\bmh_u\|^2_2 + \No}\bmh^H_u\!\left(\bmy-\sum_{j\neq u} \bmh_j z_j\right)\!.
\end{align}
\rev{This expression is exactly the CD update rule for the $u$th entry of $\bmz$. For every iteration, we can compute~\eqref{eq:CDrule} sequentially for each user $u=1,\ldots,U$, where we immediately re-use the new result $\hat{z}_u$ for the $u$th user in subsequent steps. We repeat this procedure for a total number of $K$ iterations in order to obtain an estimate for $\tilde{\bms}^\text{MMSE}=\bmz^{(K)}$, where~$\bmz^{(K)}$ is the end result of the above-described iterative process.}

\subsubsection{CD-based BOX Equalization} 
Analogously to CD-based MMSE equalization, we can derive the update rule for the  BOX equalization problem \eqref{eq:BOXproblem}. Even though the characteristic function $g^\text{BOX}(\bmz)=\chi(\bmz\in\setC_\setO)$ is not differentiable, a similar approach that uses subgradients (instead of gradients) enables one to derive the following closed-form expression \cite{Gordon12}:
\begin{align} \label{eq:CDruleBOX}
 \hat{z}_u = \mathrm{proj}_{\setC_{\setO}}\!\left( \frac{1}{\|\bmh_u\|^2_2}\bmh^H_u\!\left(\bmy-\sum_{j\neq u} \bmh_j z_j\right)\!\right)\!.
\end{align}
Here, $ \mathrm{proj}_{\setC_\setO}(\cdot)$ is the orthogonal projection onto the convex polytope $\setC_\setO$ and is given by 
\begin{align} \label{eq:projection}
  \mathrm{proj}_{\setC_\setO}(w ) = \left\{\begin{array}{ll}
  w & \text{if } w\in\setC_\setO \\
  \argmin_{q\in\setC_\setO} |w-q| & \text{if } w\notin\setC_\setO.
  \end{array}\right.
\end{align}
In words, if the argument $w\in\mathbb{C}$ is within the set $\setC_\setO$, then the projection  outputs $w$; if $w$ is outside the set $\setC_\setO$, the projection outputs the value $q$ that is closest to $w$ within the set~$\setC_\setO$ in terms of the Euclidean distance. 
We emphasize that for many practically-relevant constellation sets $\setO$, the projection~\eqref{eq:projection} can be carried out efficiently. For any QAM constellation, for example, we independently clip the real and imaginary part of $w$ onto the interval $[-\alpha,+\alpha]$, \rev{where $\alpha$ is the radius of the tightest box that covers the QAM constellation (see \secref{sec:boxdetectorexact} for the details).} For BPSK with~$\setO=\{-1,+1\}$, we clip the real part of $w$ onto the  interval $[-1,+1]$ and set the imaginary part to zero.\footnote{\rev{Orthogonal projections for PSK constellations sets are also possible. The development of efficient algorithms for PSK systems is left for future work.}}

\subsection{Optimized Coordinate Descent (OCD)}

Instead of blindly computing the updates \eqref{eq:CDrule} and \eqref{eq:CDruleBOX} for MMSE and BOX equalization, respectively, we perform preprocessing and algorithm restructuring in order to minimize the amount of (recurrent) operations during each of the $k=1,\ldots,K$ iterations. \rev{These optimizations entail \emph{no} performance loss, i.e., both methods, OCD and CD, deliver exactly the same results.}
We refer to the resulting method as the \underline{o}ptimized \underline{CD} algorithm (short OCD), which is summarized in \algref{alg:OCD}.
 OCD supports both BOX and MMSE equalization and the individual optimization steps are as follows.\footnote{The OCD algorithm proposed in the conference version of this paper~\cite{Wu2016} differs from the one presented here. The operations in OCD as proposed here have been restructured in order to (i) support MMSE as well as BOX equalization and (ii) reduce the hardware complexity.}

\begin{algorithm}[t]
\caption{Optimized Coordinate Descent (OCD)\label{alg:OCD}}
\begin{algorithmic}[1]
\STATE \textbf{\em inputs:} ${\bmy}$,  $\bH$, and $N_0$
\STATE \textbf{\em initialization:} 
\STATE \quad $\bmr = \bmy$ and ${\bmz}^{(0)} = \bZero^{U\times 1}$
\STATE \quad {MMSE mode:} $\alpha=\No$ and $\setC=\mathbb{C}$
\STATE \quad  {BOX mode:} $\alpha=0$ and $\setC=\setC_\setO$
\STATE \textbf{\em preprocessing:} 
\STATE \quad $d_u^{-1} = (\|\bmh_u\|_2^2+\alpha)^{-1}$, $u=1,\ldots,U$
\STATE \quad  $p_u = d_u^{-1}\|\bmh_u\|_2^2$, $u=1,\ldots,U$
\STATE \textbf{\em equalization:} 
\FOR{$k = 1,\ldots, K$}
\FOR{$u = 1,\ldots, \MT$}
\STATE $ z_u^{(k)} = \mathrm{proj}_\setC\!\left(d_u^{-1}\bmh^H_u\bmr + p_uz^{(k-1)}_u\right)$  \label{eq:alg1}
\STATE $\Delta z_u^{(k)} = z_u^{(k)} - z_u^{(k-1)} $  \label{eq:alg2}
\STATE $\bmr \gets  \bmr - \bmh_u\Delta z_u^{(k)}$  \label{eq:alg3}
\ENDFOR
\ENDFOR
\STATE \textbf{\em outputs:} $\tilde{\bms}=[{z}^{(K)}_1,\ldots,{z}^{(K)}_\MT]^T$
\end{algorithmic}
\end{algorithm}

\subsubsection{Preprocessing}
To reduce the computational complexity, OCD precomputes certain key quantities that can be re-used during each of the $k=1,\ldots,K$ iterations. This preprocessing step not only results in significant complexity savings during the iterative process \rev{(compared to CD)}, but also simplifies our hardware  implementation (see \secref{sec:impl}).
In particular, we precompute so-called (regularized) inverse squared column norms of~$\bH$, i.e., $d_{u}^{-1} = (\|\bmh_u\|_2^2+\alpha)^{-1}$ for $u=1,\ldots,U$, with $\alpha\geq0$, as well as regularized gains $p_{u} = d_u^{-1}\|\bmh_u\|_2^2$ for $u=1,\ldots,U$.  In MMSE mode, the regularization parameter is given by $\alpha=\No$; in BOX mode,  the regularization parameter is given by $\alpha=0$, which yields $p_{u} =1$, $u=1,\ldots,U$. 

\begin{figure*}[tp]
\centering
\subfigure[32 BS antennas and 8 users.]{\includegraphics[width=0.32\textwidth]{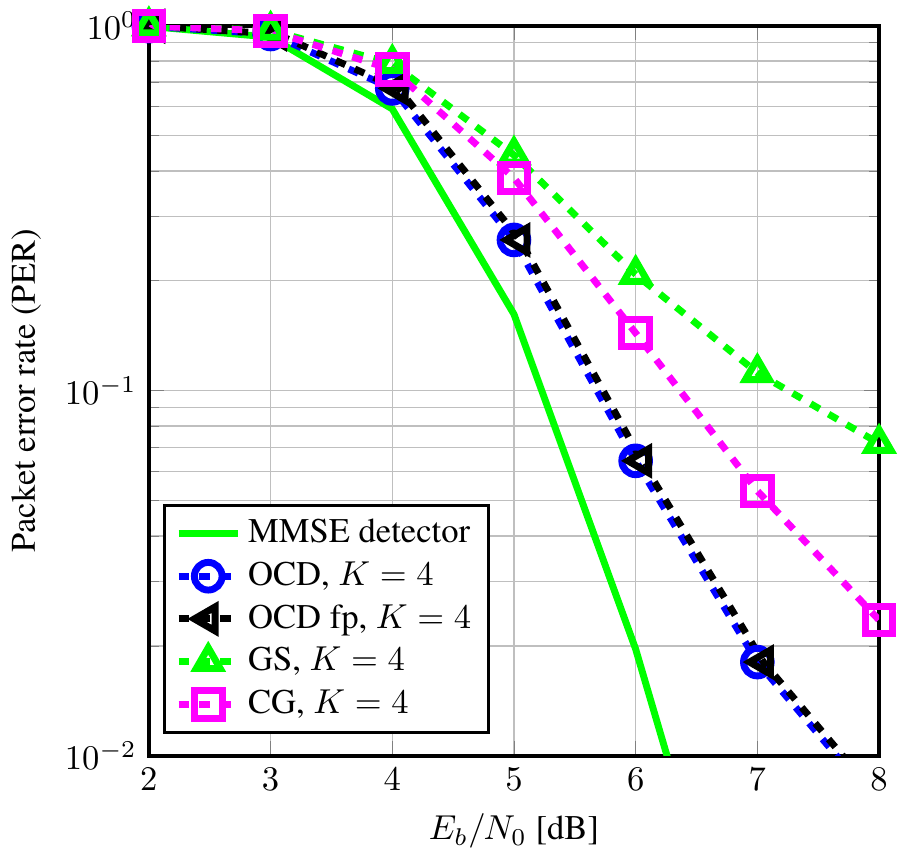}\label{fig:32x8_bler}}
\subfigure[64 BS antennas and 8 users.]{\includegraphics[width=0.32\textwidth]{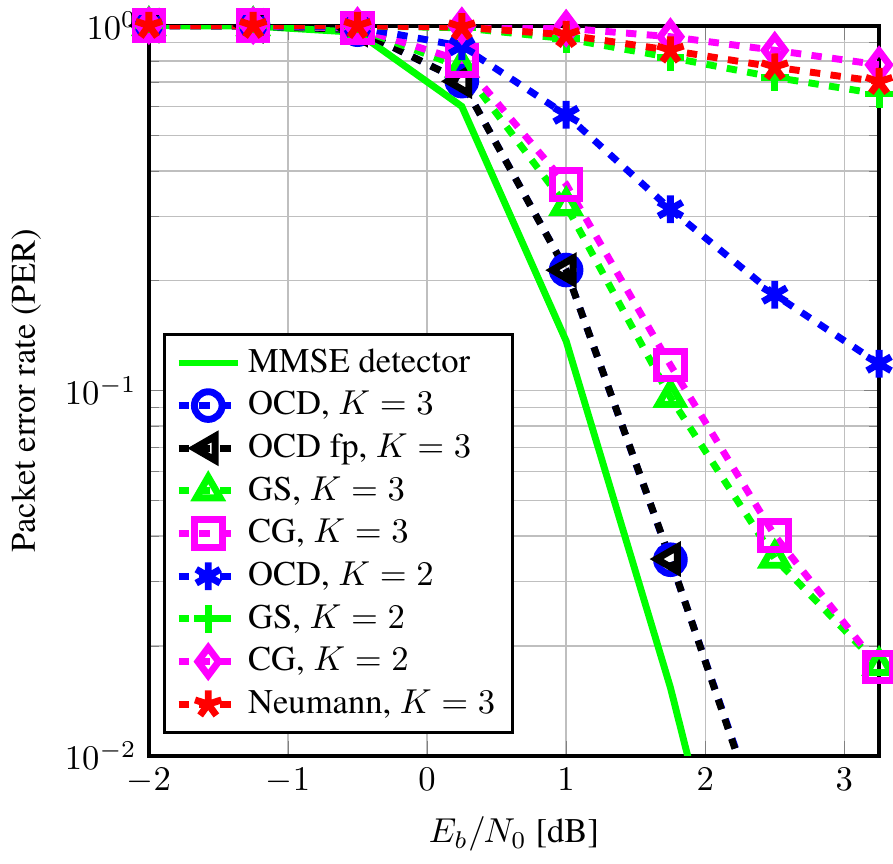}\label{fig:64x8_bler}}
\subfigure[128 BS antennas and 8 users.]{\includegraphics[width=0.32\textwidth]{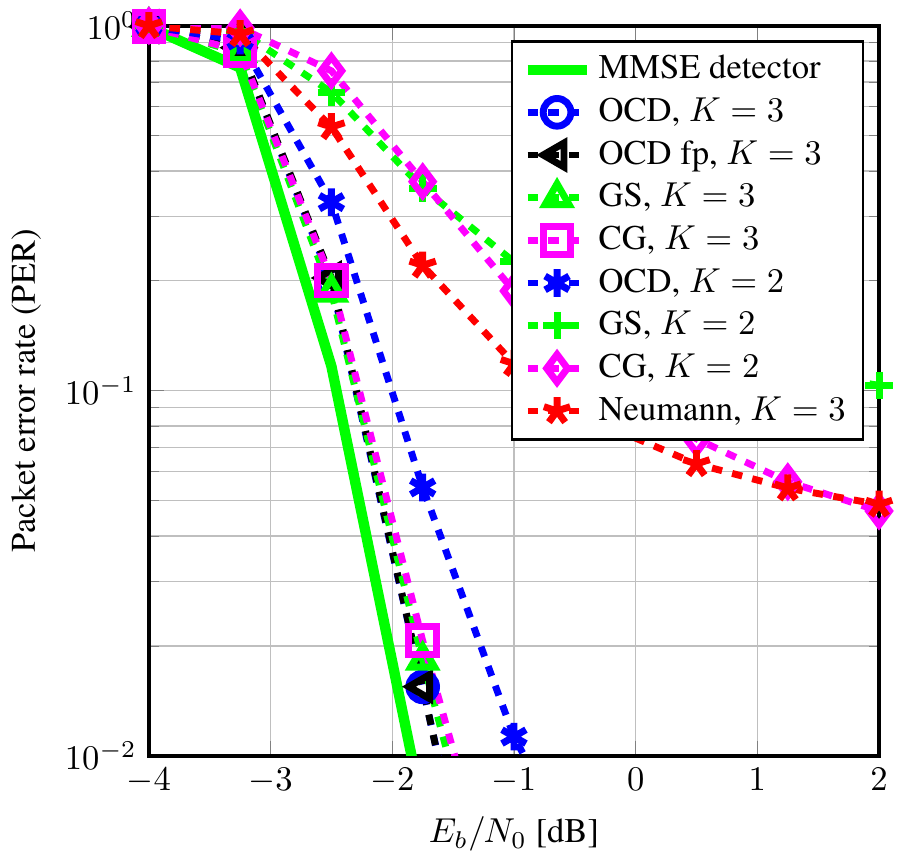}\label{fig:128x8_bler}}
\caption{Packet error rate (PER) for a massive MU-MIMO-OFDM system (``fp'' denotes fixed-point performance). Optimized coordinate descent (OCD) with box-constrained equalization achieves close-to-MMSE PER performance and outperforms the other three approximate equalization methods~\cite{Yin2015,Wu2014,WZXXY2016}.} \label{fig:michaelwusfavoritereference}
\end{figure*}

\begin{figure*}[tp]
\centering
\subfigure[32 BS antennas and 8 users.]{\includegraphics[width=0.32\textwidth]{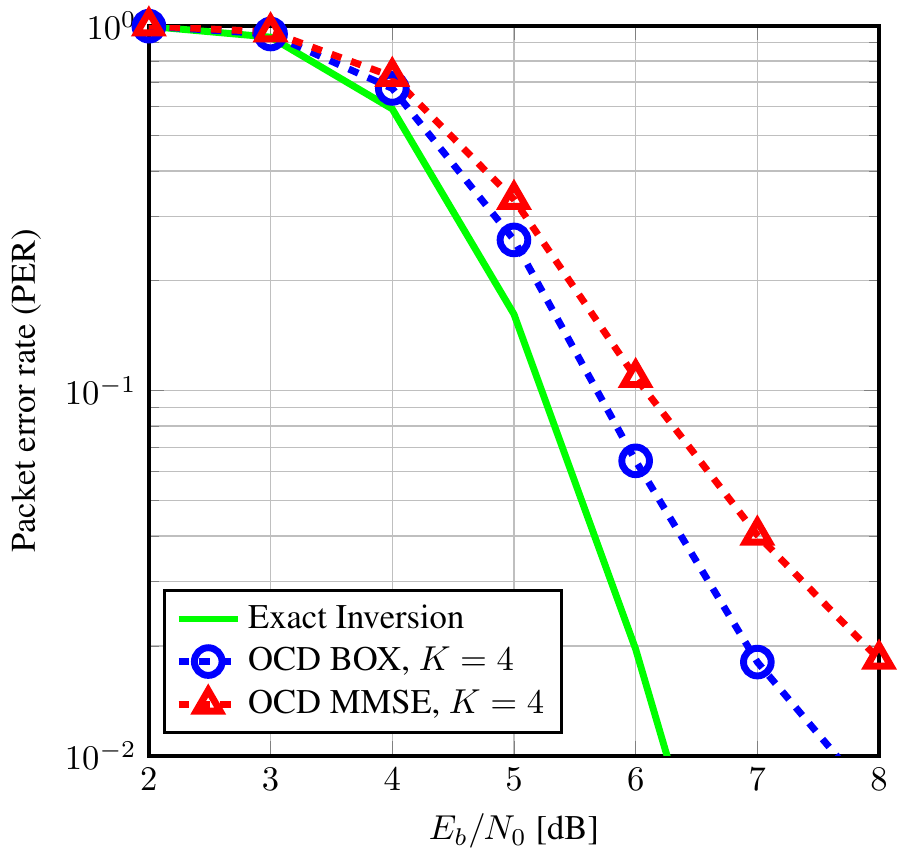}\label{fig:BLER32x8CDONOFF}}
\subfigure[64 BS antennas and 8 users.]{\includegraphics[width=0.32\textwidth]{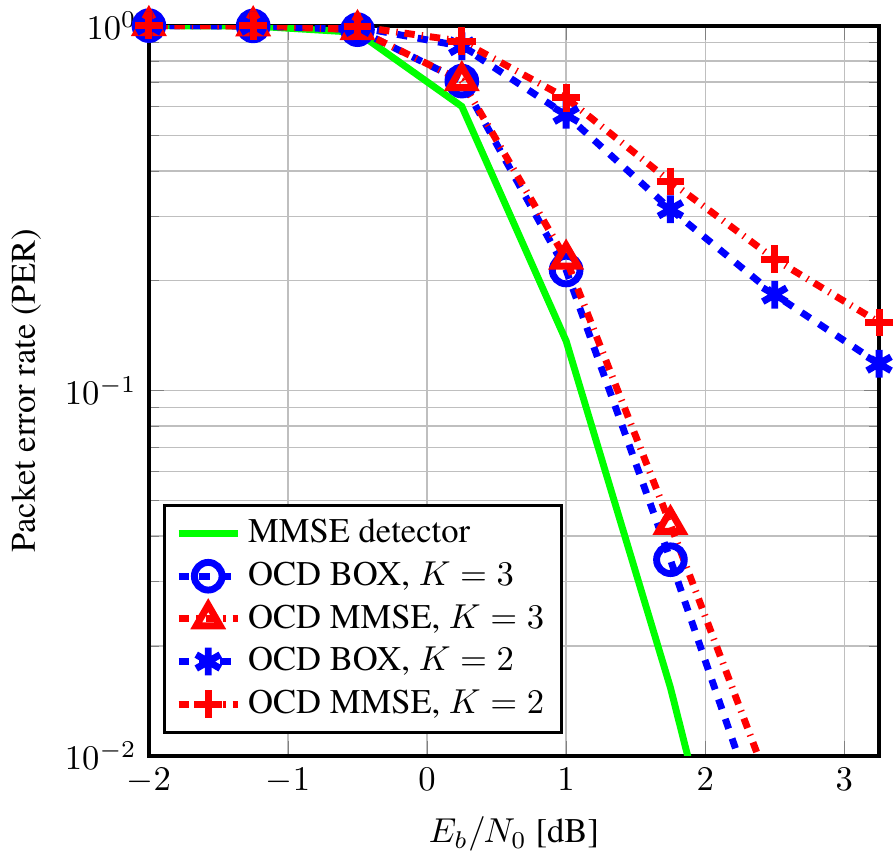}\label{fig:BLER64x8CDONOFF}}
\subfigure[128 BS antennas and 8 users.]{\includegraphics[width=0.32\textwidth]{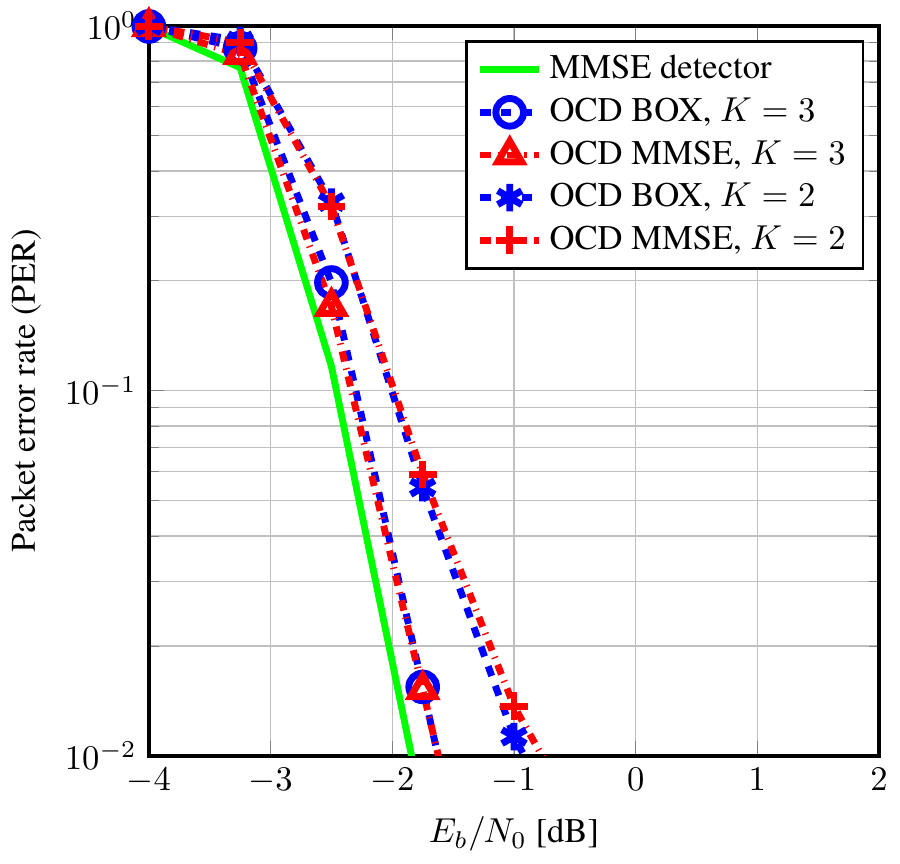}\label{fig:BLER128x8CDONOFF}}
\caption{Packet error rate (PER) for a massive MU-MIMO-OFDM system. BOX equalization outperforms MMSE equalization, especially for systems with a smaller BS-to-user antenna ratio. Furthermore, both approximate equalization methods achieve near-exact MMSE performance for a small number of iterations.} \label{fig:michaelwusfavoritereference2}
\end{figure*}

\subsubsection{Equalization}
In order to avoid recurrent operations during the equalization process, OCD performs incremental updates and re-uses intermediate quantities during each of the $k=1,\ldots,K$ iterations.
In essence, we perform sequential updates on the so-called residual approximation vector, which is defined as  
\begin{align} \label{eq:residual}
\bmr=\bmy - \sum_{j=1}^\MT \bmh_j z_j^{(k)}
\end{align}
at every algorithm iteration $k=1,\ldots,K$ and for each user $u=1,\ldots,\MT$. Note, however, that we do \emph{not} recompute this residual approximate vector for every iteration and user from scratch. In contrary, we update the residual approximation vector in every iteration and for each user by first computing the symbol estimates $z_u^{(k)}$ on line~\ref{eq:alg1} of  \algref{alg:OCD}. We then compute a so-called delta value $\Delta z_u^{(k)}$ on line~\ref{eq:alg2}, which enables us to update the residual~$\bmr$ on line~\ref{eq:alg3} without calculating the residual \eqref{eq:residual} explicitly. 

\rev{As mentioned above, OCD delivers exactly the same results as CD, but does so at significantly lower computational complexity.
In fact, the original CD algorithm in \secref{sec:CDalgo} requires one complex-valued inner product  and $\MT-1$ complex scalar-by-vector multiplications per iteration~$k$, whereas the proposed OCD algorithm requires only one inner product and one complex scalar-by-vector multiplication. More precisely, for MMSE equalization, CD requires $4BU^2+2U$ real-valued multiplications\footnote{\rev{We count $4$ real-valued multiplications per complex-valued multiplication.}} per iteration $k$, whereas OCD requires only $8BU+4U$ real-valued multiplications. Hence, for a large number of BS antennas $B$, OCD requires roughly $U/2$ times lower complexity than CD per iteration.}

\subsection{LLR Approximation for OCD}
\label{sec:llrapprox}
\rev{To compute the LLR values~\eqref{eq:exactinversellr} for MMSE and BOX equalization using OCD, we must resort to an approximation as we never explicitly compute the inverse~$\bA_w^{-1}$. To this end, we use the approximation put forward in \cite{Yincg,YWWDCS14b} for SC-FDMA-based systems. For OFDM, this approach simplifies significantly and corresponds to approximating the channel gains by  
$\tilde{\mu}_{w,i} =  d_{w,i}^{-1}g_{w,i}$,
where $d_{w,i}^{-1}$ is the $i$th regularized inverse squared column norm of~$\bH_w$ and $g_{i,w}$ is the entry in the $i$th main diagonal of the Gram matrix~$\bG_w$ at subcarrier~$w$.  
Furthermore, the approach from \cite{Yincg,YWWDCS14b} applied to OFDM systems results in the following SINR approximation:
$ \tilde{\rho}_{w,i} = \tilde{\mu}_{w,i}/(1-\tilde{\mu}_{w,i})$. 
We refer the interested reader to \cite{Yincg} for more details. 
As we will show next, this LLR approximation enables near-optimal performance in massive MU-MIMO systems with  large BS-to-user-antenna ratios.}

\subsection{Error-Rate Performance}
\label{sec:simulations}

In order to assess the error-rate performance for the proposed OCD-BOX algorithm, we perform Monte-Carlo simulations in a coded 20\,MHz MIMO-OFDM uplink system with $2048$ subcarriers, where $1200$ are used for data transmission as in \rev{LTE Advanced (LTE-A)}~\cite{3GPPLTEA}. We use 64-QAM with Gray mapping and a rate-3/4 turbo code. To account for spatial and frequency correlation, we generate channel matrices using the WINNER-Phase-2 model~\cite{winner2} with $7.8$\,cm antenna spacing as in \cite{Yincg,YWWDCS14b}.  For channel decoding, we use a \rev{log-MAP} turbo decoder.
We report the packet error-rate, which is obtained by coding over one OFDM symbol with 1200 data subcarriers. \rev{The signal-to-noise-ratio (SNR) per bit in decibels, defined as} 
\begin{align}
\rev{10\log_{10}\!\left(\frac{\Eb}{\No}\right)=10\log_{10}\!\left(\frac{\Ex{\|\bms\|^2}}{Q\Ex{\|\bmn\|^2}} \right)\!.}
\end{align}

\rev{Figures \ref{fig:michaelwusfavoritereference} and \ref{fig:michaelwusfavoritereference2} compare the packet error rate (PER) for OCD-BOX with other exact and approximate data-detection methods for massive MU-MIMO systems with various antenna configurations.} In particular, we show PER results for Neumann-series detection~\cite{Wu2014}, CG-based detection~\cite{Yin2015}, and Gauss-Seidel (GS)-based detection~\cite{WZXXY2016}. We also include an exact linear MMSE equalizer as a reference. 
For all considered antenna configurations,  OCD-BOX outperforms Neumann, CG, and GS detection for the same iteration count. 
\rev{We see that OCD with BOX equalization (OCD-BOX for short) achieves near-exact MMSE performance for only three iterations ($K=3$) for $64$ and $128$ BS antennas, whereas $K=4$ is required for the ``not-so-large'' system with $32$ BS antennas; lower values of $K$ result in a high error floor. These results confirm that for larger BS-to-user-antenna ratios, approximate linear data detectors approach the performance of the MMSE detector. We note that for the considered antenna configurations, linear MMSE detection achieves near-ML performance~\cite{Wu2014}.}

\rev{Figures \ref{fig:BLER32x8CDONOFF}, \ref{fig:BLER64x8CDONOFF}, and \ref{fig:BLER128x8CDONOFF} compare the PER  for OCD-BOX against OCD with MMSE equalization (short OCD-MMSE). The performance of OCD-BOX is superior than that of OCD-MMSE, especially in the $32$ BS antenna, $8$ user case. In general, the performance difference is more pronounced for smaller BS-to-user-antenna ratios.} \rev{This observation is in accordance to recent theoretical results~\cite{JMS2016conf}, and can be addressed to the fact that the box constraint around the constellation is tighter than the quadratic penalty $g^\text{MMSE}(\bmz)=\No\|\bmz\|_2^2$ imposed by MMSE equalization.}

\rev{We conclude by noting that for many modern wireless communication standards (such as LTE-A 
\cite{3GPPLTEA}) achieving a target PER of $10$\% is sufficient. The proposed OCD detector is able to meet this target performance  at only a small SNR loss compared to the exact MMSE-based data detector.}


\begin{figure*}[t]
\begin{center}\subfigure[OCD preprocessing mode.]{\includegraphics[width=0.7\textwidth]{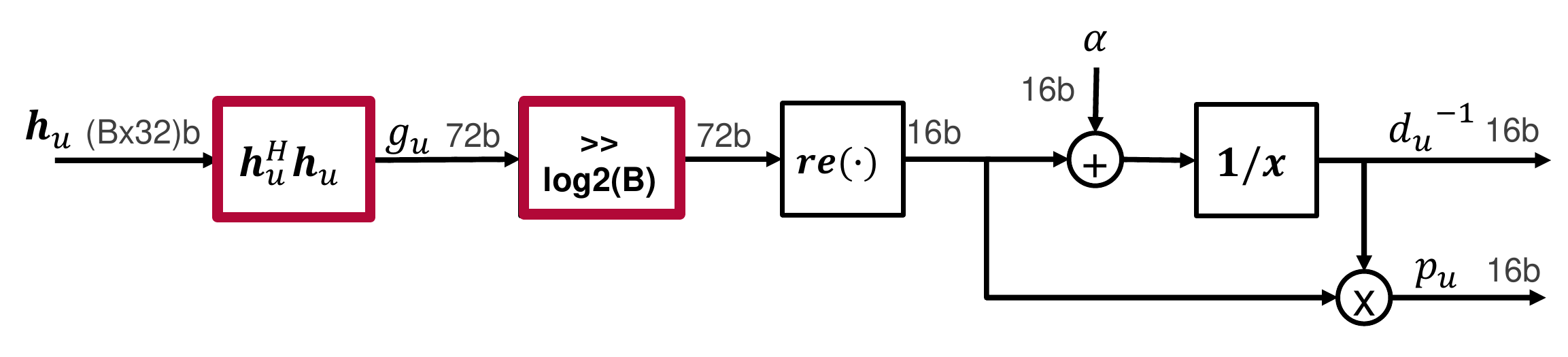}}\end{center}
\begin{center}\subfigure[OCD iteration mode.]{\centering\includegraphics[width=0.7\textwidth]{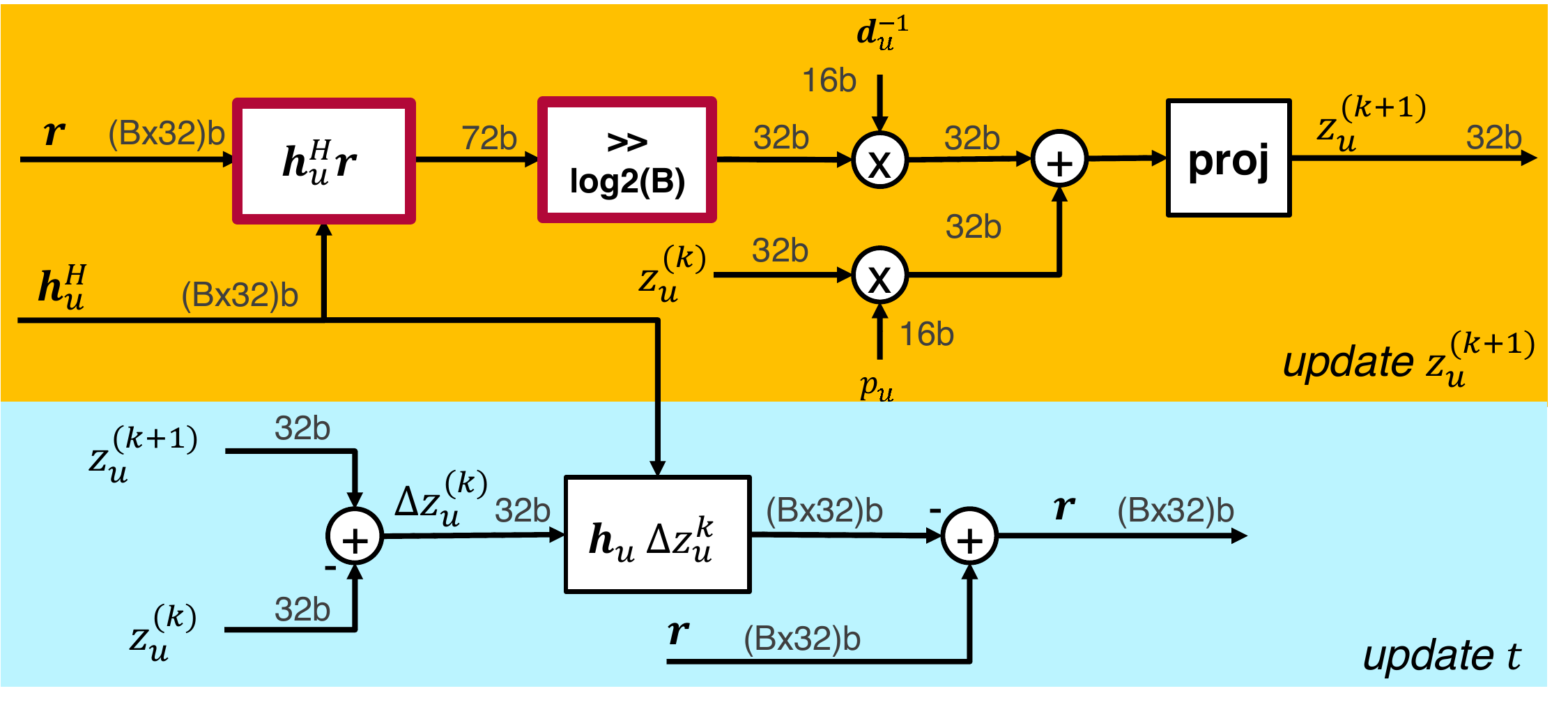}}\end{center}
\caption{High-level block diagram of the proposed OCD-based preprocessing and equalization pipeline. \rev{The pipeline is reconfigurable for various BS-antenna configurations at design time,} and is able to  perform preprocessing as well as MMSE or BOX equalization. The shared computation units between preprocessing and equalization are highlighted in red.}
\label{fig:array}
\end{figure*}

\section{VLSI Architecture}
\label{sec:impl}
We now detail our VLSI architecture for OCD-based MMSE and BOX equalization. 
The architecture was designed and optimized using Xilinx Vivado HLS~(version 2015.2), which allows us to conveniently simulate, parameterize, and generate different OCD designs that support various antenna configurations at design time.
\rev{At run-time, the proposed designs can be configured in terms of the numbers of supported users~$U$ and maximum number of iterations $K$.}

\subsection{Architecture Overview}

\figref{fig:array} shows two high-level block diagrams of the proposed OCD architecture. The inputs of our architecture are the channel matrix~$\bH_w$, the  residual error vector~$\bmr$ (which is initialized to the received vector $\bmy_w$), and the regularization parameter $\alpha$, which we initialized to $N_0$ and $0$ for MMSE and BOX equalization, respectively. 
Our architecture supports two operation modes: (a) preprocessing (lines~6--8 of \algref{alg:OCD}) and (b) OCD-based qualization~(lines 10--16). Preprocessing and equalization are carried out in a $\MR$-wide vector pipeline, i.e., we process $\MR$-dimensional vectors at a time.
In the preprocessing mode, we compute the regularized inverse squared column norms $d^{-1}_u$, $u=1,\ldots,\MT$, as well as the regularized gains $p_u$, $u=1,\ldots,\MT$. In the equalization  mode, we perform the iterations on  lines~12--13 of \algref{alg:OCD}.
In order to support these two operation modes without the need of redundant computation units, the processing pipeline shares the key building blocks used in both modes. In particular, \rev{both of the supported modes share the inner-product unit and the right-shift unit (highlighted in red in \figref{fig:array}).} The inner product unit consists of $\MR$ parallel complex-valued multipliers followed by a balanced adder tree.  %
We use multiplexers at the input of the inner product unit, which enables us to switch between preprocessing and equalization on a per-clock cycle~basis. 

One of the main implementation challenges of the proposed OCD algorithm are data dependencies between successive iterations, which prevent traditional architecture pipelining. 
In particular, as it can be seen on line 14 of \algref{alg:OCD}, each OCD iteration updates the temporary vector $\bmr$ and the vector~$\bmz_u^{(k+1)}$ given the previous vectors~$\bmr$ and~$\bmz_u^{(k)}$. 
Hence, in order to achieve high throughput, we deploy \emph{pipeline interleaving}~\cite{kaeslin2008digital}, i.e., we simultaneously process multiple subcarriers in a parallel and  interleaved manner within the same architecture. 
For example, after performing an OCD iteration for the first subcarrier, we start an  OCD iteration for the second subcarrier in the next clock cycle; we repeat this interleaving process until all pipeline stages are fully occupied. 
\rev{Our final architecture uses a total number of $24$ pipeline stages, which enables our design to achieve up to $260$\,MHz in a Xilinx Virtex-7 FPGA (see \secref{sec:implementation} for more details). We note that it is possible to achieve even higher clock frequencies by increasing the number of pipeline stages (especially for smaller small~$\MR$); this approach, however, results in a significant hardware overhead.}

\subsection{Architecture and Fixed-point Optimization}
In order to optimize the hardware efficiency of our architecture, we use fixed-point arithmetic throughout our design. 
We achieved a negligible implementation loss with $16$\,bit precision with 11 fractional bit for most internal signals; see~\figref{fig:michaelwusfavoritereference} for the fixed-point (fp) performance. Our design has an implementation loss of less than $0.2$~dB SNR \rev{(measured at a target PER of 10\%)} compared to floating-point performance for the considered scenarios, which is a result of the following two optimizations. 

\subsubsection{Inner-product unit}
This unit first computes entry-wise products of two $\MR$-dimensional vectors and then, generates the final sum of these products. We use a balanced adder tree to compute the final sum and $36$\,bit adders  to achieve sufficiently high arithmetic internal  precision. 
During preprocessing, the inner-product unit computes $\|\bmh_u\|^2_2$ (line 7 of \algref{alg:OCD}); during equalization, the same unit computes $\bmh_u^H(\bmr)$~(line 12). As both of these terms are close to $\MR$ (for large values of $\MR$), we shift these terms by $b=\ceil{\log_2(\MR)}$\,bits to the right in order to reduce the dynamic range. 
Since we shift $\|\bmh_u\|^2_2$ by $b$ to the right, when we compute the reciprocal value, $d_u^{-1} = (\|\bmh_u\|_2^2+\alpha)^{-1}$, we effectively shift the reciprocal value~$d_u^{-1}$ by $b$\,bits to the left. 
In the inner-product unit, we also shift the term $\bmh_u^H\bmr$ by $b$\,bits to the right. 
Consequently, we do not need to undo both of these shifts, as they cancel out during the multiplication on line 12 of \algref{alg:OCD}.

\subsubsection{Reciprocal unit}
This unit consists of two parts. The first part normalizes the input value to the range $[0.5,1]$, which is accomplished using a leading-zero detector and programmable shift to the left. The second part generates a reciprocal value for the normalized input using a look-up table (LUT). We use a FPGA BRAM18 to implement a $18$\,bit, 2048 entry LUT, where the leading $11$\,bits of the normalized input value are used to point to the entry in the LUT that stores the associated normalized reciprocal.
Finally, we denormalize the normalized reciprocal value by another left shift.

\begin{table}[tp]
\centering
\caption{Implementation results on a Xilinx Virtex-7 XC7VX690T~FPGA~for different BS antenna numbers}
\begin{tabular}{lllll}
\toprule
{Array size} & ${\MR=32}$ &${\MR=64}$ & ${\MR=128}$ \tabularnewline
\midrule
\rev{\# of  Slices}    & 2\,873   &   6\,508   & 11\,094\tabularnewline
\rev{\# of LUTs}     & 6\,059   & 12\,588   & 23\,914\tabularnewline
\rev{\# of FFs}       & 10\,704 & 24\,801  &  43\,008\tabularnewline
\rev{\# of DSP48s} & 198       & 390         & 774      \tabularnewline
\rev{\# of BRAM18s} & 2          &  2            & 2         \tabularnewline
\midrule
Max.\ clock frequency & 261\,MHz & 261\,MHz & 258\,MHz\tabularnewline
\bottomrule
\end{tabular}
\label{tbl:implresultcd}
\end{table}

\begin{table*}[tp]
\centering
{\color{black}\caption{Area breakdown on a Xilinx Virtex-7 XC7VX690T~FPGA~for different BS antenna numbers}\label{tbl:areabreakdown}
\begin{tabular}{lllllll}
\toprule
& Main units                               & \# of Slices &  \# of LUTs &  \# of FFs &  \# of DSP48s &  \# of BRAM18s \tabularnewline
\midrule
\parbox[t]{2mm}{\multirow{5}{*}{\rotatebox[origin=c]{90}{$B=32$}}} 
& {$\bmr$ update unit}                & 256 (8.91\%)     &  1\,024 (16.9\%)    & 0  (0\%)               & 0   (0\%)      & 0     (0\%) \tabularnewline
& {Inner-product unit}                  &  811 (28.2\%)    & 1\,045 (17.3\%)     & 2\,416 (22.6\%)   & 96 (48.5\%)    & 0  (0\%) \tabularnewline
& {$ \bmh_u\Delta z_u$ scaling unit } & 265   (9.22\%)   &  249 (4.11\%)        & 1\,137 (10.6\%)   & 96 (48.5\%)    & 0 (0\%) \tabularnewline
& {Miscellaneous }                       & 1\,541 (53.6\%)  & 3\,741 (61.74\%)  & 7\,151 (66.8\%)  & 6  (3.0\%)      & 2 (100\%)  \tabularnewline
& {Total}                                       & 2\,873 (100\%)  &  6\,059 (100\%)     & 10\,704  (100\%) & 198 (100\%)  & 2 (100\%)   \tabularnewline
\midrule
\parbox[t]{2mm}{\multirow{5}{*}{\rotatebox[origin=c]{90}{$B=64$}}} 
& {$\bmr$ update unit}                & 512 (7.87\%)      &  2\,048 (16.3\%)    & 0  (0\%)               & 0   (0\%)      & 0     (0\%) \tabularnewline
& {Inner-product unit}                  & 1\,627 (25.0\%)  & 2\,006 (15.9\%)     & 5\,776 (23.3\%)   & 192 (49.2\%)    & 0  (0\%) \tabularnewline
& {$ \bmh_u\Delta z_u$ scaling unit} & 485   (7.45\%)    &  505 (4.01\%)        & 2\,161 (8.71\%)   & 192 (49.2\%)    & 0 (0\%) \tabularnewline
& {Miscellaneous }                       & 3\,884 (59.7\%)  & 8\,029 (63.8\%)     & 16\,864 (68.0\%) & 6  (1.6\%)      & 2 (100\%)  \tabularnewline
& {Total}                                       & 6\,508 (100\%)   &  12\,588 (100\%)   & 24\,801  (100\%) & 390 (100\%)  & 2 (100\%)   \tabularnewline
\midrule
\parbox[t]{2mm}{\multirow{5}{*}{\rotatebox[origin=c]{90}{$B=128$}}} 
& {$\bmr$ update unit}                & 1\,024 (9.23\%)     &  4\,096 (17.1\%)    & 0  (0\%)               & 0   (0\%)      & 0     (0\%) \tabularnewline
& {Inner-product unit}                  &  3\,447 (31.1\%)    & 4\,109 (17.2\%)     & 11\,676 (27.0\%)   & 384 (49.6\%)    & 0  (0\%) \tabularnewline
& {$ \bmh_u\Delta z_u$ scaling unit} & 1\,955   (17.6\%)   &  5\,120 (21.4\%)        & 4\,211 (9.72\%)   & 384 (49.6\%)    & 0 (0\%) \tabularnewline
& {Miscellaneous }                       & 4\,668 (42.1\%)  & 10\,589 (44.3\%)  & 27\,421 (63.3\%)  & 6  (0.8\%)      & 2 (100\%)  \tabularnewline
& {Total}                                       & 11\,094 (100\%)  &  23\,914 (100\%)     & 43\,308  (100\%) & 774 (100\%)  & 2 (100\%)   \tabularnewline
\bottomrule
\end{tabular}}

\end{table*}

\begin{table}[tp]
\centering
\caption{Throughput and latency on a Xilinx Virtex-7 XC7VX690T~FPGA for $K$~iterations and $64$-QAM, and $128$ BS and $8$  user antennas}
\begin{tabular}{llllll}
\toprule
                                 & ${K=1}$ &${K=2}$ & ${K=3}$ & ${K=4}$\tabularnewline
\midrule
\rev{Max.\ throughput [Mb/s]}    & $1\,363$  &     $496$ & $376$ & $302$\tabularnewline
Latency [$\upmu$s] & 1.58 & 2.33 & 3.08 & 3.82 \tabularnewline
\bottomrule
\end{tabular}
\label{tbl:implresultcd_throughput}
\end{table}

\section{Implementation Results and Comparison}
\label{sec:implementation}

We now show FPGA implementation results and compare our design to the recently proposed data-detectors for massive MU-MIMO systems in~\cite{Yin2015,Wu2014,WZXXY2016,CGS2016a}.

 \begin{table*}[t]
\caption{Comparison of  $128\times8$ data detectors for massive MU-MIMO~system on a Xilinx Virtex-7 XC7VX690T FPGA }
\label{tbl:implcomp}
	\begin{minipage}[c]{2\columnwidth}
	\centering
  	\begin{tabular}{lllllll}
  		\toprule
  		{Detector} & CG~\cite{Yin2015} & Neumann~\cite{Wu2014} & Gauss-Seidel~\cite{WZXXY2016} & TASER~\cite{CGS2016a} & {OCD} 
		\tabularnewline
		\midrule
		Performance & near-MMSE & near-MMSE & near-MMSE & near-ML & near-MMSE\tabularnewline
		Highest modulation & 64-QAM & 64-QAM & 64-QAM & QPSK & 64-QAM \tabularnewline
		Iteration count $K$ & 3 & 3 & 1\footnote{The method uses a special Neumann-series initializer followed by one GS iteration.} & 3 &3\tabularnewline
  		\midrule
  		{\# of slices}& 1\,094 (1.0\%) & 48\,244 (45\%) & n.a. & 4\,350 (4.0\%) & 11\,094 (10\%) \tabularnewline
  		{\# of LUTs}& 3\,324 (0.8\%) & 148\,797 (34\%) & 18\,976 (4.3\%) & 13\,779 (3.2\%) & 23\,914 (5.5\%) \tabularnewline
  		{\# of FFs} & 3\,878 (0.4\%) &   161\,934 (19\%) & 15\,864 (1.8\%) & 6\,857 (0.8\%) & 43\,008 (4.96\%)  \tabularnewline
  		{\# of DSP48s} & 33 (0.9\%)& 1\,016  (28\%) & 232 (6.3\%) & 168 (5.7\%) & 774 (21.5\%) \tabularnewline
  		{\# of \rev{BRAM18s}} & 1          &  16            & 6  & 0 & 2         \tabularnewline
  		\midrule
  		Maximum clock frequency [MHz] & 412& 317 & 309 & 225 & 258 \tabularnewline
  		{Latency [clock cycles]} & 951 & 196 & n.a. & 72 & 795  \tabularnewline
  		{Maximum throughput} [Mb/s]& 20 &  621  & 48 & 50 & \rev{376} \tabularnewline
  		\midrule
  		{Throughput/LUTs}  & {6\,017} & {4\,173} & 2\,530 & 3\,629 &  15\,597   \tabularnewline
  		\bottomrule
  	\end{tabular}
  	 \end{minipage}  	
  \end{table*}

\subsection{FPGA Implementation Results}
We designed three different implementations for the following BS antenna configurations:  $\MR=32$, \mbox{$\MR=64$} and $\MR=128$.
For each configuration, we provide post place-and-route implementation results  on a Xilinx \mbox{Virtex-7} XC7VX690T FPGA. All our designs support $U\leq32$ users and $K\leq256$ OCD iterations; both of these parameters can be set at run-time.

\rev{The hardware complexity, resource utilization, and maximum clock frequency results are summarized in \tblref{tbl:implresultcd}. 
We note that there is no particular critical path in all our designs as Vivado HLS evenly optimizes the delays among all pipeline stages. 
A detailed area breakdown of the main units is shown in \tblref{tbl:areabreakdown}. The ``$\bmr$ update unit'' corresponds to the output adder in \figref{fig:array}(b); the ``inner-product unit'' corresponds to the unit that computes $\bmh^H_u\bmh_u$ and $\bmh^H_u\bmr$ in \figref{fig:array}(a) and  \figref{fig:array}(b), respectively; the ``$\bmh_u\Delta z_u$ scaling unit'' corresponds to the scaling block in \figref{fig:array}(a); all remaining circuitry has been flattened by Vivado HLS and is subsumed in ``miscellaneous.''}
Since the proposed architecture performs operations on $\MR$-dimensional vectors, the resource utilization (excluding the BRAMs) scales linearly with~$\MR$. Since the quantities $\bH_w$ and~$\bmy_w$ are assumed to be stored in external memories, our OCD architecture only uses two BRAM18s: one for the reciprocal LUT and one to store the regularized channel gains~$p_u$, $u=1,\ldots,\MT$.

The maximum achievable throughput as well as the processing latency are shown in \tblref{tbl:implresultcd_throughput}. We see that the throughput only depends on the maximum iteration number~$K$ and the clock frequency, but does not depend on $\MT$. The reason is because the number of bits per subcarrier and the number of clock cycles required to process $24$ subcarriers grows linearly with respect to $\MT$. For example, doubling $\MT$ doubles the number of bits per subcarrier. However, since the number of OCD updates is~$KU$, the number of required clock cycles also doubles; this results in a constant throughput. For $K=3$ iterations, \rev{which was shown in~\figref{fig:michaelwusfavoritereference} to achieve} near-optimal performance, our design achieves $376$\,Mb/s. Hence, the use of only three parallel instances (to process subcarriers in parallel) would easily exceed $1.1$\,Gb/s, while consuming less than \rev{$65$\% of the FPGA's BRAM18s }(cf.~\tblref{tbl:implcomp}). 

The processing latency increases roughly linearly with respect to $K$ and $\MT$. More specifically, the processing latency of this design is approximately $24(K+1)U+O$ clock cycles, where $O$ is the number of cycles required to flush the pipeline. Typically, $26$ cycles are required to flush the pipeline, the exact value of $O$ depends on $\MR$.  The (approximately) linear increase in $K$ can be seen in  \tblref{tbl:implresultcd_throughput} and for $K=3$, our design requires only $3.08$\,$\upmu$s to produce its first equalized output.

\subsection{Comparison}
  
\tblref{tbl:implcomp} compares OCD to other, recently proposed large-scale MIMO data detectors, namely the conjugate gradient (CG)-based detector \cite{Yin2015}, the Neumann-series detector \cite{Wu2014}, the Gauss-Seidel (GS) detector~\cite{WZXXY2016}, and triangular approximate semidefinite relaxation (TASER)~\cite{CGS2016a}. All of these detectors have been implemented on the same FPGA and for a 128 BS antenna, 8 user system. 
We see that  for the same system configuration, OCD outperforms all other designs in terms of hardware efficiency, which we define as  throughput per FPGA LUTs. 
Furthermore, our OCD detector achieves superior PER performance than the CG, Neumann, and GS detector (see Figs.~\ref{fig:64x8_bler} and \ref{fig:128x8_bler}), which demonstrates the effectiveness of OCD. TASER, in contrast, achieves better error-rate performance for the considered antenna configuration\footnote{TASER achieves near-ML performance in ``not-so-massive'' MIMO systems, where the number of users is comparable to the number of BS antennas.} but only supports QPSK constellations. 
\rev{We note that the throughput of (approximate) linear detectors, such as the ones in \cite{Yin2015,Wu2014,WZXXY2016} scales linearly in the number of bits $Q$ per symbol; for TASER, however, the throughput is limited by QPSK modulation, which prevents this detector to achieve comparable throughputs as the other approximate methods.}

\rev{In summary, we see that OCD outperforms the next-best design (namely the CG-detector from \cite{Yin2015}) by more than $2.6\times$ in terms of hardware efficiency. The reasons for this advantage are due to the facts that (i) OCD can be implemented in a very regular and parallel manner and (ii) preprocessing requires significantly lower complexity compared to that of the other detectors that require the computation of the regularized Gram matrix $\bA_w$, which can be a significant burden in massive MU-MIMO-OFDM systems.}


\section{Conclusions}
\label{sec:conclusions}

We have proposed a novel coordinate descent (CD)-based data detector, called optimized CD (OCD), for massive MU-MIMO systems that use orthogonal frequency division multiplexing (OFDM). 
The proposed OCD detector enables high-performance linear MMSE and non-linear box-constrained data detection using a simple, parallel VLSI architecture that requires low hardware complexity.
Our FPGA reference design achieves $376$\,Mb/s for a $128$ BS antenna, $8$ user system, and substantially outperforms existing  approximate linear data-detection methods in terms of hardware efficiency and/or error-rate performance.
Our results show that OCD enables realistic OFDM-based massive MU-MIMO systems to support tens of users communicating with hundreds of BS antennas, while achieving high throughput at low implementation costs.

There are many avenues for future work. OCD can also be used for linear and non-linear precoding in the massive MU-MIMO downlink; a corresponding study is part of ongoing work. Computing exact soft-output values for OCD-based detection (for MMSE and BOX equalization) is an interesting open research problem. Finally, accelerated CD algorithms have been proposed recently~\cite{lee2013efficient}; such methods may lead to even faster convergence and hence, could enable higher throughput at the same error-rate performance when implemented in VLSI. 


%
%
\section{Acknowledgments}
C. Studer would like to thank Tom Goldstein, Charles Jeon, Shahriar Shahabuddin for insightful discussions on the box-constrained equalization method. 
The work of M. Wu and J. R. Cavallaro was supported in part by Xilinx Inc., and by the US National Science Foundation (NSF) under grants ECCS-1408370, CNS-1265332, and ECCS-1232274. The work of C.~Studer was supported in part by Xilinx Inc.\ and by the US NSF under grants ECCS-1408006 and CCF-1535897.


\balance


\end{document}